\documentclass[12pt]{article}
\usepackage{latexsym}
\usepackage{graphicx}
\def\bseq{\begin{subequation}}  
\def\eseq{\end{subequation}}
\def\bsea{\begin{subeqnarray}}  
\def\esea{\end{subeqnarray}}

\newcommand{\bbox}{\lower.2ex\hbox{$\Box$}}

\newcommand{\beq}{\begin{equation}}
\newcommand{\eeq}{\end{equation}}
\newcommand{\bea}{\begin{eqnarray}}
\newcommand{\eea}{\end{eqnarray}}
\newcommand{\ena}{\end{eqnarray}}

\renewcommand{\a}{\alpha}
\renewcommand{\b}{\beta}

\renewcommand{\d}{\delta}

\newcommand{\pa}{\partial}
\newcommand{\g}{\gamma}
\newcommand{\G}{\Gamma}

\newcommand{\e}{\epsilon}

\renewcommand{\L}{\Lambda}
\newcommand{\m}{\mu}

\newcommand{\n}{\nu}

\newcommand{\p}{\pi}

\newcommand{\s}{\sigma}

\newcommand{\ad}{{\dot{\alpha}}}
\newcommand{\bd}{{\dot{\beta}}}

\begin{document}
\begin{titlepage}
\begin{flushright}
IFUM-FT-668\\

\end{flushright}
\vspace{1cm}
\begin{center}
{\Large \bf The one-loop effective action of noncommutative\\
\vspace{3mm}

  ${\cal N}
=4$ super Yang-Mills is gauge invariant}
\vfill
{\large \bf
Mario Pernici~, Alberto Santambrogio ~and~
Daniela Zanon}\\
\vskip 7mm
{\small
Dipartimento di Fisica dell'Universit\`a di Milano
and\\ INFN, Sezione di Milano, Via Celoria 16,
20133 Milano, Italy\\}
\end{center}
\vfill
\begin{center}
{\bf Abstract}
\end{center}
{\small We study the gauge transformation of the recently
computed one-loop four-point function
of ${\cal N}=4$
supersymmetric Yang-Mills theory with gauge group $U(N)$.
The contributions from nonplanar diagrams are not
gauge invariant. We compute their gauge variation and show that it
is  cancelled
by the variation from  corresponding terms of the one-loop five-point
function. This mechanism is general: it insures
the gauge invariance of the noncommutative one-loop effective action.}
\vspace{2mm} \vfill \hrule width 3.cm
\begin{flushleft}
e-mail: mario.pernici@mi.infn.it\\
e-mail: alberto.santambrogio@mi.infn.it\\
e-mail: daniela.zanon@mi.infn.it
\end{flushleft}
\end{titlepage}

Recently noncommutative gauge field theories have attracted much
attention \cite{all,SW,others}. The motivations for such an interest
reside primarily on two basic facts: on one side they
represent the field theory limit of open
strings in the presence of a constant $B$-field  \cite{string},
on the other side
they offer new examples of non local field theories whose properties
can be studied in a rather simple setting  \cite{perturb}.
Supersymmetry can be added into the game and in fact the analysis can be
performed using a full superspace formulation
\cite{FL,susy,schweda,DZ}.

In \cite{DZ} it has been shown that the quantization of
noncommutative supersymmetric Yang-Mills theories can be studied within
a superfield framework that exactly parallels its commutative
counterpart. In particular one can take advantage of the background
field method which drastically simplifies perturbative calculations
\cite{GS,GZ}.
Indeed in \cite{DZ,ASDZ} quantum corrections to the noncommutative
${\cal N}=4$ supersymmetric Yang-Mills classical action have been
considered. It has been shown that at one loop the noncommutative
$\b$-function is zero, since as it happens
 in the commutative case \cite{GS2}, the two- and
three-point contributions to the effective action vanish. Moreover
the one-loop four-point function has been computed.
The result has been compared with the off-shell
continuation of the corresponding result from string amplitudes
\cite{*trek} and complete agreement has been checked \cite{ASDZ}.

The one-loop four-point function is given by the sum of terms that
arise from planar and nonplanar diagrams. The background field
formalism introduces the external fields through covariant
objects with a multiplication in terms of the
$*$-product. After integration on the loop momentum the contributions
from planar graphs are still expressible with the $*$-product,
while the contributions from nonplanar graphs require the introduction of
generalized multiplication rules, i.e. $*'$- and $*_3$-products
\cite{*trek,gar,wise}.
As pointed out in
\cite{*trek,DZ,ASDZ} the net result is that while the planar sector
respects the gauge invariance of the classical action,
the non planar sector is not gauge invariant.

It becomes relevant to investigate whether the Ward identities are
satisfied in noncommutative gauge theories.
The standard reasoning in commutative gauge theories is that, if the
regularized tree-level
action is background gauge invariant (or BRS invariant) then the
effective action
satisfies the background Ward identities (or the Slavnov-Taylor identities).
This argument relies on properties like the invariance of the path-integral
measure under
background gauge (or BRS) transformations
which are well established in commutative gauge theories.

In this letter we address the issue of gauge invariance for the
effective action of ${\cal N}=4$ super Yang-Mills.
We compute the gauge variation
of the nonplanar terms in the one-loop four-point function presented
in \cite{ASDZ}
and show that it is  cancelled
by the variation from  corresponding terms of the one-loop five-point
function. We discuss the cancellation mechanism in general and argue
that it insures
the gauge invariance of the noncommutative one-loop effective action.
First we set the notation and briefly review the formalism and the
results obtained in \cite{DZ,ASDZ}.
\vspace{0.8cm}

The noncommutative
${\cal N}=4$ supersymmetric Yang-Mills classical action
written in terms of ${\cal N}=1$ superfields (we use the
notations and conventions adopted in \cite{superspace}) is given by
\bea
&&S= \frac{1}{g^2}~{\rm Tr} \left( \int~ d^4x~d^4\theta~ e^{-V}
\bar{\Phi}_i e^{V} \Phi^i +\frac{1}{2} \int ~d^4x~d^2\theta~ W^2
+\frac{1}{2} \int ~d^4x~d^2\bar{\theta}~ \bar{W}^2 \right.\nonumber\\
&&\left.\left.~~~~~~~~~~~~~+\frac{1}{3!} \int ~d^4x~d^2\theta~ i\e_{ijk}
 \Phi^i
[\Phi^j,\Phi^k] + \frac{1}{3!}\int ~d^4x~d^2\bar{\theta}~ i\e^{ijk} \bar{\Phi}_i
[\bar{\Phi}_j,\bar{\Phi}_k] \right)\right|_*
\label{SYMaction}
\eea
where the $\Phi^i$ with $i=1,2,3$ are three chiral
superfields, and the $W^\a= i\bar{D}^2(e_*^{-V}*D^\a e_*^V)$ are the gauge
superfield strengths. All the fields are Lie-algebra valued, e.g.
$\Phi^i=\Phi^i_a T^a$, in the adjoint representation of $U(N)$.
In (\ref{SYMaction}) the symbol $|_*$ indicates that the
superfields are multiplied using  the $*$-product defined as
\beq
(\phi_1 * \phi_2)(x,\theta,\bar{\theta})\equiv e^{\frac{i}{2}
\Theta^{\m\n}\frac{\pa}{\pa x^\m}\frac{\pa}{\pa y^\n}}~
\phi_1(x,\theta,\bar{\theta})\phi_2(y,\theta,\bar{\theta})|_{y=x}
\label{starprod}
\eeq
The theory can be quantized using the background field method
\cite{DZ}, which essentially consists in a non linear splitting between
a quantum prepotential $V$ and a background superfield, via covariant
derivatives
\bea
&&\nabla_\a = e_*^{-V}* D_\a~ e_*^{V}
~\rightarrow~e_*^{-V}*\nabla^B_\a ~ e_*^{V} \nonumber\\
&&~~~~~\nonumber\\
&&\bar{\nabla}_\ad = ~\bar{D}_\ad
~\rightarrow~
\bar{\nabla}^B_\ad
\label{backcovder}
\eea
On the r.h.s. of (\ref{backcovder})
the covariant derivatives are expressed in terms of
background connections, i.e.
\beq
\nabla^B_\a= D_\a-i{\bf{\G}}_\a \qquad \qquad \bar{\nabla}^B_\ad= \bar{D}_\ad-
i\bar{\bf{\G}}_\ad \qquad\qquad \nabla^B_a=\pa_a-i{\bf{\G}}_a
\label{backcovderconn}
\eeq
In this way the external background enters in the
quantum action implicitly in the background covariant derivatives through the
connections and explicitly in the background field strength
${\bf W}_\a=\frac{i}{2}[\bar{\nabla}^{B\ad},\{\bar{\nabla}^B_\ad,\nabla^B_\a\}]$.

After the splitting the classical action is invariant under two
separate sets of gauge transformations.\\
The quantum transformations:
\bea
&&e^V~\rightarrow~e_*^{i\bar{\L}}* e^V* e_*^{-i\L}\qquad\qquad\qquad
\Phi^i~\rightarrow~e_*^{i\L}\Phi^i e_*^{-i\L}\nonumber\\
&&~~~~~~\nonumber\\
&&\nabla_\a~\rightarrow~e_*^{i\L}\nabla_\a
e_*^{-i\L}\qquad\qquad\quad~
\qquad
\bar{\nabla}_\ad
~\rightarrow~\bar{\nabla}_\ad \nonumber\\
&&~~~~~~\nonumber\\
&&\nabla^B_\a~\rightarrow~\nabla^B_\a\qquad\qquad\qquad\quad\quad~
\qquad
\bar{\nabla}^B_\ad
~\rightarrow~\bar{\nabla}^B_\ad
\label{quantumvar}
\eea
with supergauge parameter $\L$ which is a
covariantly chiral superfield, $\bar{\nabla}^B_\ad \L=0$.\\
The background transformations:
\bea
&&e^V~\rightarrow~e_*^{iK}* e^V* e_*^{-iK}\qquad\qquad\qquad
\Phi^i~\rightarrow~e_*^{iK}\Phi^i e_*^{-iK}\nonumber\\
&&~~~~~~\nonumber\\
&&\nabla_\a~\rightarrow~e_*^{iK}\nabla_\a
e_*^{-iK}\qquad\qquad\quad~
\qquad
\bar{\nabla}_\ad
~\rightarrow~e_*^{iK}\bar{\nabla}_\ad e_*^{-iK}\nonumber\\
&&~~~~~~\nonumber\\
&&\nabla^B_\a~\rightarrow~e_*^{iK}\nabla^B_\a e_*^{-iK}\qquad\qquad
\qquad\quad~
\bar{\nabla}^B_\ad
~\rightarrow~e_*^{iK}\bar{\nabla}^B_\ad  e_*^{-iK}
\label{backvar}
\eea
with a real superfield parameter, $K=\bar{K}$. The quantum gauge
invariance needs gauge-fixing and this can be done in a background
covariant way
adding to the classical Lagrangian background covariantly
chiral gauge-fixing functions, $\nabla_B^2 V$ and $\bar{\nabla}_B^2 V$.
As emphasized in \cite{DZ,ASDZ} the quantization for the
noncommutative theory proceeds following the same steps as for the
commutative case. At one loop, amplitudes with external vector
fields receive contributions from quantum $V$ fields only,
since the chiral matter loops are exactly cancelled by the ghost loops
because of statistics.

The quantum quadratic action, relevant for the
computation of the one-loop vector corrections is
\bea
&&-\frac{1}{2g^2} {\rm Tr}\left( ~V*\frac{1}{2} \pa^a\pa_a V
-iV*[{\bf \G}^a ,\pa_aV]_*
-\frac{1}{2}V*[{\bf\G}^a ,[{\bf
\G}_a,V]_*]_* \right.\nonumber\\
&&~~~~~~~~~~~~~~~~~~~ -iV*\{ {\bf W}^\a , D_\a V\}_*
-V*\{ {\bf W}^\a,[{\bf\G}_\a ,V]_*\}_*\nonumber\\
&&~~~~~~~~~~~~~~~~~~~ \left.-iV*\{
 \bar{{\bf W}}^\ad, \bar{D}_\ad V\}_*-V*
 \{ \bar{{\bf W}}^\ad,[\bar{{\bf \G}}_\ad , V]_*\}_*\right)
\label{oneloopaction}
\eea
The background connections have been defined
in (\ref{backcovderconn}) and we have introduced the
notation $[A,B]_*=A*B-B*A$.
The interactions with
the background fields are at most linear in the $D$ spinor derivatives.
Therefore one can immediately conclude that the first non vanishing correction
to the effective action is given by the four-point function
since $D$-algebra supergraph rules require at least two $D$'s and
two $\bar{D}$'s in the loop. The complete result has been obtained in
\cite{ASDZ} and we make reference to that work for details of the
calculation. Here we briefly explain how the various contributions
are assembled and we present the final answer.
\vspace{0.8cm}

The vertices that enter in the evaluation of the four-point function
are
\beq
\frac{i}{2g^2}{\rm Tr}\left( V*\{ {\bf W}^\a , D_\a V\}_*+V*\{
 \bar{{\bf W}}^\ad, \bar{D}_\ad V\}_*\right)
 \label{relvert}
 \eeq
When inserted in the loop, depending on the order in which the quantum $V$'s
are Wick contracted, they produce two types of terms,
an untwisted $P$-term and a twisted $T$-term \cite{DZ}
\bea
&&P\rightarrow {\rm Tr}(T^aT^bT^c)~ e^{-\frac{i}{2}(k_1
\times k_2+k_2\times k_3+k_1\times k_3)}={\rm Tr}(T^aT^bT^c)~
 e^{-\frac{i}{2}k_2\times k_3}\nonumber\\
&&~~~~~\nonumber\\
&&T\rightarrow -{\rm Tr}(T^cT^bT^a)~
e^{\frac{i}{2}(k_1
\times k_2+k_2\times k_3+k_1\times k_3)}=-{\rm Tr}(T^cT^bT^a)~
e^{\frac{i}{2}k_2\times k_3}
\label{PandT}
\eea
where we have defined
 $k_i\times k_j\equiv
(k_i)_\m\Theta^{\m\n}(k_j)_\n$ and used $k_1+k_2+k_3=0$.

 The external background fields are
arranged as follows
\bea
&&{\cal T}(1a,2b,3c,4d)={\bf W}^{\a a}(p_1){\bf W}_\a^b(p_2)
\bar{{\bf W}}^{\ad c}(p_3)
\bar{{\bf W}}_\ad^d(p_4)~~~~~~~~~~~~~~~~~~~~~~~~\nonumber\\
&&~~~~~~~~~~~~~~~~~~~~~~~+{\bf W}^{\a a}(p_1)\bar{{\bf W}}^{\ad b}(p_2)
\bar{{\bf W}}_\ad^c(p_3)
{\bf W}_\a^d(p_4)\nonumber\\
&&~~~~~~~~~~~~~~~~~~~~~~~-{\bf W}^{\a a}(p_1)\bar{{\bf W}}^{\ad b}(p_2)
{\bf W}_\a^c(p_3)
\bar{{\bf W}}_\ad^d(p_4)~+~{\rm h.c.}
\label{symmetric}
\eea
The above expression, completely symmetric in the exchanges of any
couple $(1a)\leftrightarrow (2b)\leftrightarrow (3c) \leftrightarrow
(4d)$, can be compared easily with the bosonic result from string amplitude
calculations since we have
\bea
&&\int d^2\theta ~d^2\bar{\theta}~\left[{\bf W}^{\a a}(p_1){\bf W}_\a^b(p_2)
\bar{{\bf W}}^{\ad c}(p_3)
\bar{{\bf W}}_\ad^d(p_4)\right.~~~~~~~~~~~~~~~~~~~~~~~~\nonumber\\
&&~~~~~~~~~~~~~~~~~~~~~~~+{\bf W}^{\a a}(p_1)\bar{{\bf W}}^{\ad b}(p_2)
\bar{{\bf W}}_\ad^c(p_3)
{\bf W}_\a^d(p_4)\nonumber\\
&&~~~~~~~~~~~~~~~~~~~~~~~\left.-{\bf W}^{\a a}(p_1)\bar{{\bf W}}^{\ad b}(p_2)
{\bf W}_\a^c(p_3)
\bar{{\bf W}}_\ad^d(p_4)~+~{\rm h.c.}\right]\nonumber\\
&&\rightarrow t^{\a\b\g\d\m\n\rho\s} F_{\a\b}^a(p_1)F_{\g\d}^b(p_2)
F_{\m\n}^c(p_3)F_{\rho\s}^d(p_4)
\label{bosonic}
\eea
where $t^{\a\b\g\d\m\n\rho\s}$ is the symmetric tensor given
e.g. in (9.A.18) of \cite{GSW}. The one-loop four-point diagrams
contain four propagators which produce a factor
\beq
I_0(k;p_1,\dots,p_4)=\frac{1}{(k+p_1)^2 k^2 (k-p_4)^2 (k+p_1+p_2)^2}
\label{box}
\eeq
They can be divided into planar and non planar contributions.
There are two planar graphs: one with four $P$ vertices and one with
four $T$ vertices with a  trace factor  ${\rm Tr}(T^p T^q T^r T^s)$
from the $U(N)$ matrices.
Their contribution is symbolically written as
\beq
{\cal P}(p_1,\dots,p_4)=P_1P_2P_3P_4+T_1T_2T_3T_4
\label{PPPP}
\eeq
The nonplanar
diagrams contain either two twisted and two untwisted vertices
or else one
twisted and three untwisted vertices (equivalently three twisted
and one untwisted).
For the first group the trace on the
$U(N)$ matrices gives a factor like ${\rm Tr}(T^p T^q){\rm
Tr}(T^rT^s)$, while for the second group it gives ${\rm Tr}( T^p) {\rm Tr}(T^q
T^rT^s)$. Their contribution is symbolically written as
${\cal A}(k,p_1,\dots,p_4)+{\cal B}(k,p_1,\dots,p_4)$ with
\bea
&&{\cal A}(k,p_1,\dots,p_4)=P_1P_2T_3T_4+P_1T_2T_3P_4
+P_1T_2P_3T_4~~~~~~~~~~\nonumber\\
&&~~~~~~~~~~~~~~~~~~~~~~~+T_1 T_2P_3P_4+T_1P_2P_3T_4+T_1P_2T_3P_4
\label{calA}
\eea
and
\bea
&&{\cal B}(k,p_1,\dots,p_4)=P_1P_2P_3T_4+P_1T_2P_3P_4+P_1P_2T_3P_4+
T_1P_2P_3P_4~~~~~~~~\nonumber\\
&&~~~~~~~~~~~~~~~~~~~~~~~~+T_1T_2T_3P_4+T_1P_2T_3T_4+T_1T_2P_3T_4
+P_1T_2T_3T_4
\label{calB}
\eea
Using the above definitions the complete one-loop vector four-point function
is given by \cite{ASDZ}
\bea
&&\G_{\rm total}= \frac{1}{4}
\int d^2\theta~d^2\bar{\theta}~
\frac{d^4p_1~d^4p_2~d^4p_3~d^4p_4}{(2\p)^{16}}~\d(\sum
p_i)~{\cal T}(1a,2b,3c,4d)\nonumber\\
&&~~~~~~~~~\int d^4k~I_0(k;p_1,\dots,p_4)
~\left[ {\cal P}(p_1,\dots,p_4)~ +~{\cal A}(k,p_1,\dots,p_4)~
\right.\nonumber\\
&&~~~~~~~~~~~~~~~~~~~~~~~~~~~~~~~~~~~~~~~~~~~~~~~~~~~~~~~\left.
+~{\cal B}(k,p_1,\dots,p_4)~\right]
\label{total}
\eea
The above expression can be evaluated in the low-energy approximation
 $p_i\cdot p_j$ small,
with $p_i\times p_j$ finite. In this case the integration on the
$k$-loop momentum can be performed exactly. Introducing an
IR mass regulator in a background gauge invariant manner, and the definition
$p\circ p=p_\m\Theta^{\m\n}\Theta_{\rho\n}p^{\rho}$,
the complete result for the low-energy planar and nonplanar contributions
to the four-point function can be written as
\bea
&&\G_{\rm l.e.}=\frac{\p^2}{4}
\int d^2\theta~d^2\bar{\theta}~
\frac{d^4p_1~d^4p_2~d^4p_3~d^4p_4}{(2\p)^{16}}~\d(\sum p_i)~
~~~~~~~~~~~~~~~~~~~~~~~~~~~~~~~~~~~~~~~\nonumber\\
&&~~~~~~~~~~\left[{\bf W}^{\a a}(p_1){\bf W}_\a^b(p_2)
\bar{{\bf W}}^{\ad c}(p_3)
\bar{{\bf W}}_\ad^d(p_4)
+{\bf W}^{\a a}(p_1)\bar{{\bf W}}^{\ad b}(p_2)
\bar{{\bf W}}_\ad^c(p_3)
{\bf W}_\a^d(p_4)\right.\nonumber\\
&&~~~~~~~~~~~~~~~~~~~~~~~~~~~~~~~~~~\left.-{\bf W}^{\a a}(p_1)
\bar{{\bf W}}^{\ad b}(p_2){\bf W}_\a^c(p_3)
\bar{{\bf W}}_\ad^d(p_4)~+~{\rm h.c.}\right]
\nonumber\\
&&~~~~\nonumber\\
&&~~~~~~\left\{ \frac{1}{6m^4} ~{\rm Tr}(T^a T^b T^c
T^d)~e^{\frac{i}{2}(p_2\times p_1+p_1\times p_4+p_2\times p_4)}
\right.\nonumber\\
&&~~~~~~~~+~\frac{1}{m^2}~\frac{(p_1+p_2)\circ(p_1+p_2)}{2}
~\frac{\sin\left(\frac{p_1\times p_2}{2}\right)}
{\frac{p_1\times
p_2}{2}}~
\frac{\sin\left(\frac{p_3\times p_4}{2}\right)}{\frac{p_3\times
p_4}{2}}~\nonumber\\
&&~~~~\nonumber\\
&&~~~~~~~~~~~~~~~~~~~~~~~~~~~~
K_2(m\sqrt{(p_1+p_2)\circ(p_1+p_2)})
~{\rm Tr}(T^a T^b)~{\rm Tr}(T^cT^d)\nonumber\\
&&~~~~~\nonumber\\
&&~~~~~~~~-~\frac{1}{m^2}~(p_1+p_2+p_3)\circ(p_1+p_2+p_3)~
\frac{e^{-\frac{i}{2}(p_1\times p_2+p_2\times p_3+p_1\times p_3)}}
{(p_1\times p_4)(p_3\times p_4)}\nonumber\\
&&~~~~\nonumber\\
&&~~~~~~~~~~~~~~~~~~~~~~~~~~~~
K_2(m\sqrt{(p_1+p_2+p_3)\circ(p_1+p_2+p_3)})
~{\rm Tr}(T^a T^bT^c)~{\rm Tr}(T^d)
\nonumber\\
&&~~~~\nonumber\\
&&~~~~~~~~~~~~~~~~~~~~~~~~~~~~~~~~~~~~~~~~~~~~~~~~~~~~~~~~~~~
\left.+~{\rm h.c.}~\right\}
\label{lowenfinal}
\eea
Again using (\ref{bosonic}) we can extract the purely bosonic
contributions in terms of the $F_{\m\n}$'s. The resulting expression
coincides with the off-shell extrapolation of the field
theory limit obtained in \cite{*trek} from open string amplitudes.
The terms from planar diagrams can be rewritten in terms of
$*$-products between the superfield strengths, while the terms
proportional to ${\rm Tr}(T^a T^b){\rm Tr}(T^cT^d)$ and to
${\rm Tr}(T^a T^bT^c){\rm Tr}(T^d)$
corresponding to non planar diagrams reproduce the $*'$- and the
$*_3$-products respectively \cite{gar,*trek,wise}.
While the planar contributions are gauge invariant under the
background transformations in (\ref{backvar}), the non planar terms
break gauge invariance explicitly, despite the fact that they are
expressed in terms of covariant objects.
We observe that even if we
were to replace
the $*'$ and the $*_3$ operations  with
$*$-products still the nonplanar terms would not maintain gauge
invariance. The new generalized multiplication rules
are not the cause of trouble. In fact we will show
that gauge invariance is recovered
taking into account the gauge variation of corresponding terms
from the one-loop five point function.
\vspace{0.8cm}

Now we discuss in some detail the gauge variation of the one-loop
four-point function in (\ref{total}). The total variation consists in
the sum of four terms corresponding to the variation of
each external background field contained in the expression
${\cal T}(1a,2b,3c,4d)$ explicitly
given in (\ref{symmetric}). Symmetry considerations allow to repeat
the same argument for each of the four contributions, so that
in order to be specific we
consider the variation of the ${\bf W}^\a(p_1)$ superfield.
From (\ref{backvar}) we obtain the
background gauge transformation of the superfield strength
\beq
\d {\bf W}^\a= iK*{\bf W}^\a-i{\bf W}^\a*K\equiv \d_1 {\bf W}^\a + \d_2 {\bf W}^\a
\label{Wgaugevar}
\eeq
The two terms can be written in momentum space as
\bea
&&\d_1{\bf W}^\a(p_1)=i\int d^4x~ e^{-ip_1\cdot x} K*{\bf W}^\a \nonumber\\
&&~~~~~~~~~~~~~~=i\int d^4x~\int \frac{d^4p_0~d^4k}{(2\p)^8}
~e^{i(p_0+k)\cdot x}
e^{-\frac{i}{2}p_0\times k} e^{-ip_1\cdot x}K(p_0){\bf W}^\a(k)\nonumber\\
&&~~~~~~~~~~~~~~=i\int \frac{d^4p_0}{(2\p)^4}
~e^{-\frac{i}{2} p_0\times p_1} K(p_0)
{\bf W}^\a(p_1-p_0)
\label{d1}
\eea
and
\bea
&&\d_2{\bf W}^\a(p_1)=-i\int \frac{d^4p_0}{(2\p)^4}
~e^{\frac{i}{2} p_0\times p_1}
{\bf W}^\a(p_1-p_0)K(p_0)
\label{d2}
\eea
Our goal is to prove that such a variation of the
external background field ${\bf W}^\a$ in the four-point function
is compensated by the variation
of the background connection ${\bf \G}_a$ which appears as an
additional vertex from (\ref{oneloopaction}) in the corresponding
five-point function. The analysis is quite simple if it is performed
directly on the various one-loop diagrams that contribute to the
total answer.
As an example of the cancellation mechanism we focus on the
contribution in (\ref{total}) which has vertices in the
order $P_1P_2T_3T_4$ with ${\bf W}^\a(p_1){\bf W}_\a(p_2)
\bar{{\bf W}}^\ad(p_3)\bar{{\bf W}}_\ad(p_4)$ as external background.
The corresponding Feynman diagram is shown in Fig. 1.

\vskip 34pt
\noindent
\begin{minipage}{\textwidth}
\begin{center}
\includegraphics[width=0.50\textwidth]{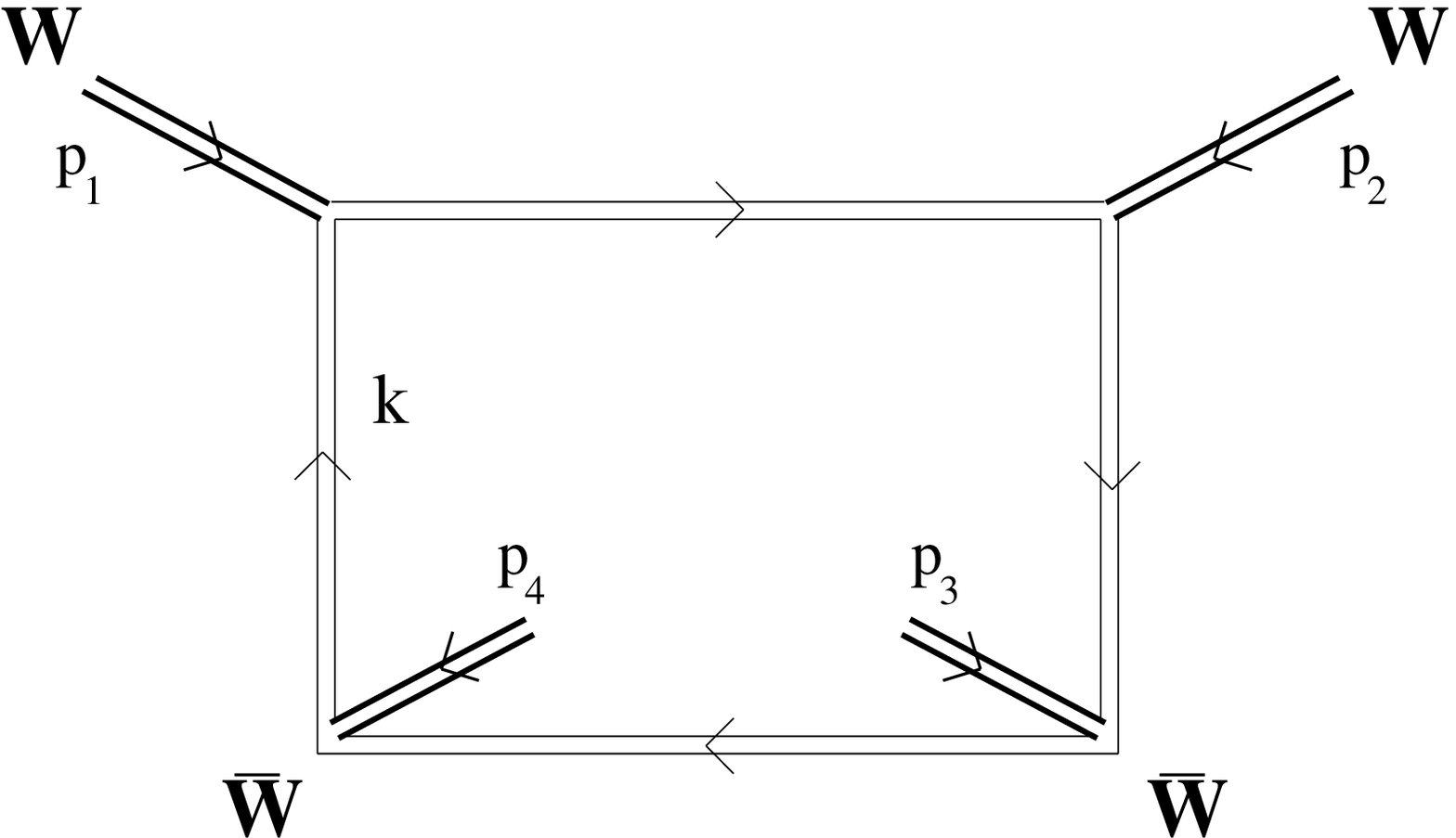}
\end{center}
\begin{center}
{\small{Figure 1:
$P_1P_2T_3T_4$ contribution to the four point function}}
\end{center}
\end{minipage}

\vskip 34pt
It gives
\bea \label{example}
&&\G_{PT}=\frac{1}{4}
\int d^2\theta~d^2\bar{\theta}~
\frac{d^4p_1~d^4p_2~d^4p_3~d^4p_4}{(2\p)^{16}}~\d(\sum
p_i)~e^{-\frac{i}{2}(p_1\times p_2-p_3\times p_4)}\\
&&~~~~~~\nonumber\\
&&~~~~{\rm Tr}[{\bf W}^\a(p_1){\bf W}_\a(p_2)]~
{\rm Tr}[\bar{{\bf W}}^\ad(p_3)\bar{{\bf W}}_\ad(p_4)]
\int d^4k~\frac{
 e^{-ik\times (p_1+p_2)}}{(k+p_1)^2 k^2 (k-p_4)^2
(k+p_1+p_2)^2}\nonumber
\eea
Now we compute its $\d_1$ gauge transformation as in (\ref{d1})
\bea
&&\d_1 \G_{PT}^{(1)}=\frac{i}{4}
\int d^2\theta~d^2\bar{\theta}~
\frac{d^4p_1~d^4p_2~d^4p_3~d^4p_4}{(2\p)^{16}}~\d(\sum
p_i)~\nonumber\\
&&~~~~~~\nonumber\\
&&~~~~~~~~~~~~~~~~~~~~
\int d^4k~\frac{e^{-\frac{i}{2}(p_1\times p_2-p_3\times p_4)}
 e^{-ik\times (p_1+p_2)}}{(k+p_1)^2 k^2 (k-p_4)^2
(k+p_1+p_2)^2}\nonumber\\
&&~~~~~~\nonumber\\
&&~~~~\int \frac{d^4p_0}{(2\p)^4}~e^{-\frac{i}{2} p_0\times p_1}
{\rm Tr}[K(p_0){\bf W}^\a(p_1-p_0){\bf W}_\a(p_2)]~
{\rm Tr}[\bar{{\bf W}}^\ad(p_3)\bar{{\bf W}}_\ad(p_4)]
\label{varexample}
\eea
The $\d_2 \G_{PT}^{(1)}$ variation is obtained in similar manner.
In the same way one compute $\d \G_{PT}^{(2)}$, $\d \G_{PT}^{(3)}$
and $\d \G_{PT}^{(4)}$ varying in (\ref{example})
${\bf W}_\a(p_2)$,
$\bar{{\bf W}}^\ad(p_3)$ and $\bar{{\bf W}}_\ad(p_4)$ respectively.
One finds that the sum  $\d \G_{PT}^{(1)}+\d \G_{PT}^{(2)}
+\d \G_{PT}^{(3)}+\d \G_{PT}^{(4)}$
is not zero. Thus we look for a
cancellation from a higher-point function.

Let us consider the contribution to the one-loop five point-function
corresponding to a supergraph with the same vertices considered above
plus the following additional vertex (cf. the action in
(\ref{oneloopaction}))
\beq
{\cal{V}}=\frac{i}{2g^2}\int~d^2\theta d^2\bar{\theta} d^4x~
{\rm Tr}V*[{\bf \G}^a ,\pa_aV]_*
\label{newvertex}
\eeq
When inserted in the loop, depending on the order in which the
quantum fields are Wick contracted it will produce contributions of
untwisted or twisted type. Now we compute its gauge variation under
the background transformations in (\ref{backvar}) and show that its
presence compensates the terms in (\ref{varexample}).
From (\ref{backvar}) we obtain the linearized background gauge variation of the
connection
\beq
\d{\bf \Gamma}_{a}=\pa_{a}K
\label{Ggaugevar}
\eeq
so that the vertex varies into
\bea
\d {\cal{V}}&=&
\frac{i}{g^2}{\rm Tr}
\int~d^2\theta d^2\bar{\theta} d^4x~
(-V*K*\Box V +\Box V*K*V)\nonumber\\
&\equiv & \d_1{\cal{V}}+\d_2{\cal{V}}
\label{5var}
\eea
As emphasized above each term in (\ref{5var}) produces an untwisted
and a twisted contribution
\beq
\d_1 {\cal{V}}\rightarrow \d_1 {\cal{V}}_P+\d_1
{\cal{V}}_T\qquad\qquad\quad \d_2 {\cal{V}}\rightarrow \d_2
{\cal{V}}_P+\d_2
{\cal{V}}_T
\eeq
We focus on ${\cal{V}}_P$. The analysis of the other terms leads
to corresponding similar conclusions.

\vskip 34pt
\noindent
\begin{minipage}{\textwidth}
\begin{center}
\includegraphics[width=0.70\textwidth]{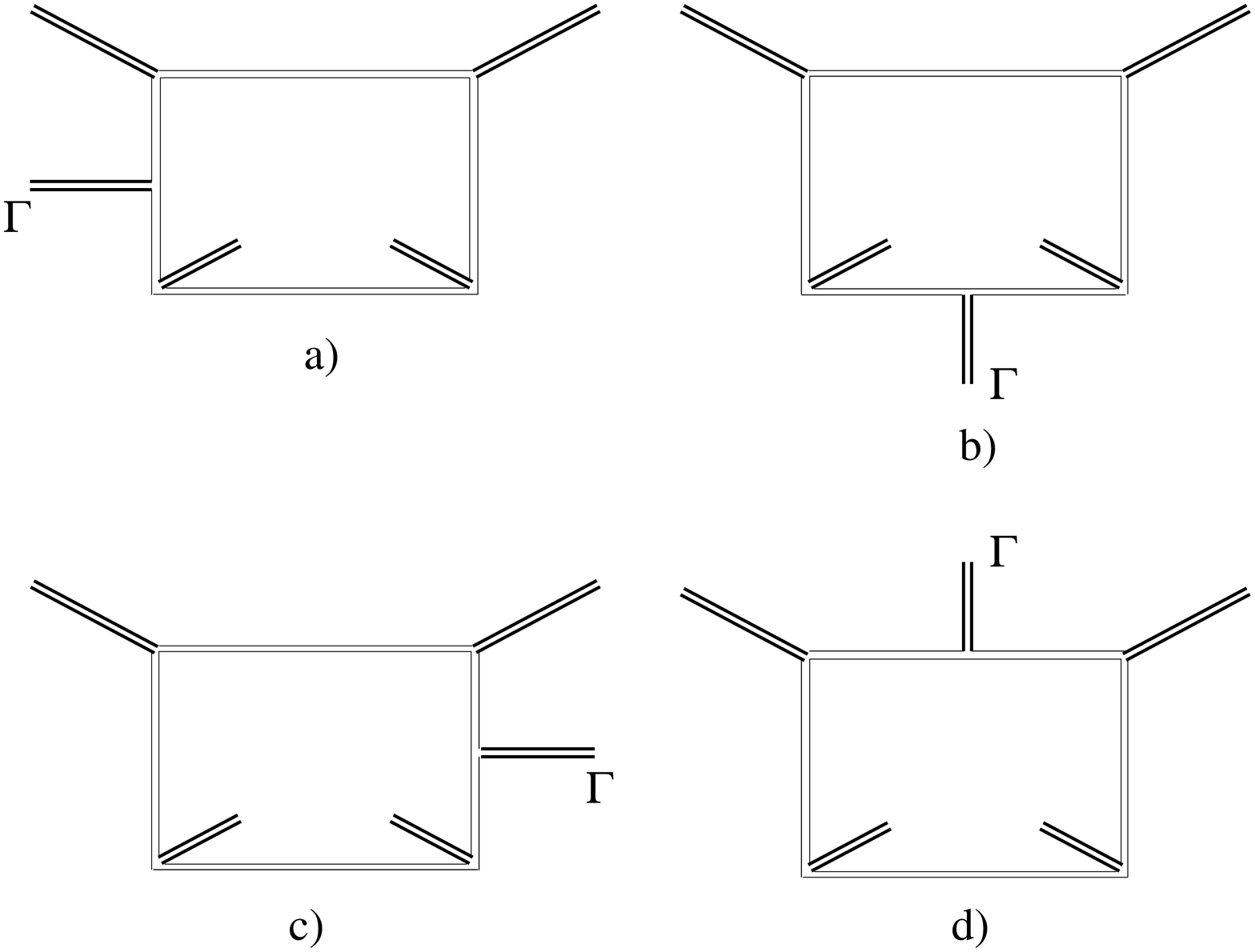}
\end{center}
\begin{center}
{\small{Figure 2:
five-point function}}
\end{center}
\end{minipage}

\vskip 34pt
At the level of the five-point function, the ${\cal{V}}$ vertex can
appear in between any of the four field strengths. The four diagrams
corresponding to the $\d {\cal{V}}_P$ variation are shown in Fig.
$2a,\dots,2d$. For each of them we have to consider $\d_1
{\cal{V}}_P$ and $\d_2{\cal{V}}_P$. Let us start with $\d_1{\cal{V}}_P$
as in Fig. $2a$.  It gives
\bea
&&\d_1{\cal{V}}_P^{(a)}\rightarrow -\frac{i}{4}\int d^2\theta~d^2\bar{\theta}~
\frac{d^4p_0~d^4p_1~d^4p_2~d^4p_3~d^4p_4}{(2\p)^{20}}~
\d(p_1-p_0+p_2+p_3+p_4+p_0)~\nonumber\\
&&~~~~~~\nonumber\\
&&~~~~~~~~~~~~~~~~
\int d^4k~\frac{e^{-\frac{i}{2}(p_1\times p_2-p_3\times p_4)}
 e^{-ik\times (p_1+p_2)}}{(k+p_1)^2(k+p_0)^2 k^2 (k-p_4)^2
(k+p_1+p_2)^2}(k+p_0)^2~e^{-\frac{i}{2} p_0\times p_1}\nonumber\\
&&~~~~~~\nonumber\\
&&~~~~~~~~~~~~~~~~~~~~~~~~~~
{\rm Tr}[K(p_0){\bf W}^\a(p_1-p_0){\bf W}_\a(p_2)]~
{\rm Tr}[\bar{{\bf W}}^\ad(p_3)\bar{{\bf W}}_\ad(p_4)]
\label{d1Ga}
\eea
It is immediate to check that this expression exactly cancels the
$\d_1$ variation in (\ref{varexample}).

\vskip 34pt
\noindent
\begin{minipage}{\textwidth}
\begin{center}
\includegraphics[width=0.80\textwidth]{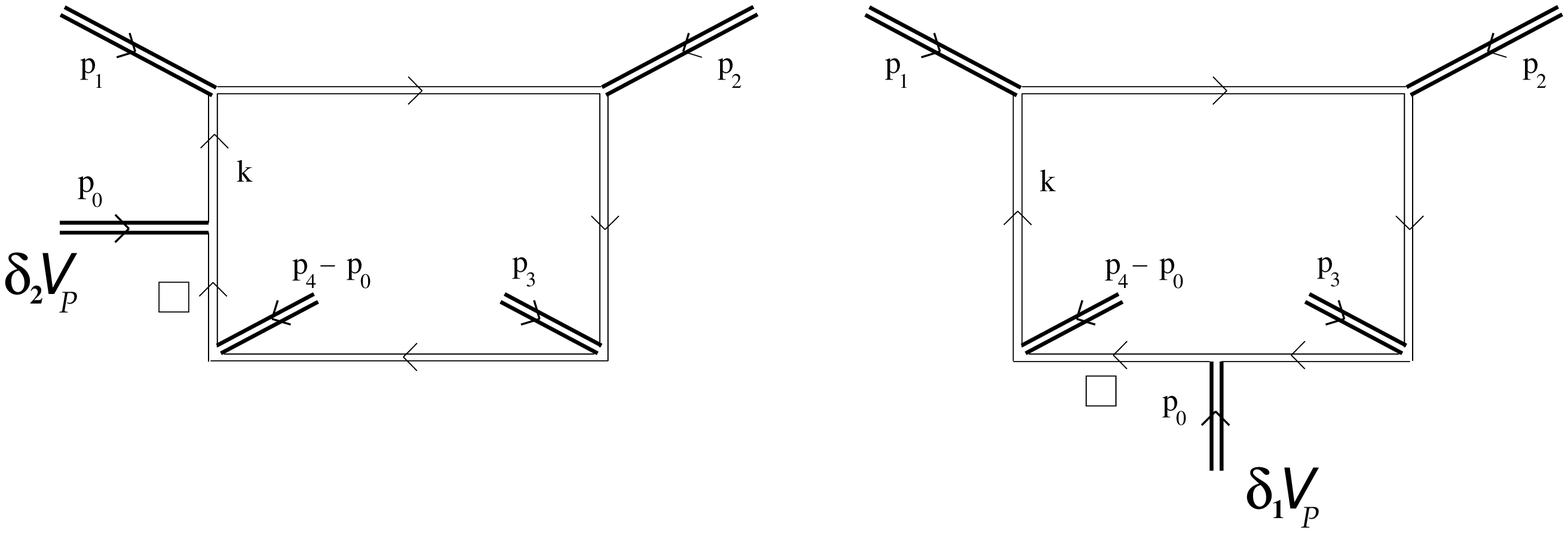}
\end{center}
\begin{center}
{\small{Figure 3:
example of cancellation}}
\end{center}
\end{minipage}

\vskip 34pt
Moreover it is rather simple  to prove that the $\d_2{\cal{V}}_P$ term as in
Fig. $2a$ compensates the terms from $\d_1{\cal{V}}_P$ as in Fig. $2b$.
Indeed with the labeling of momenta  shown in Fig. $3$ one obtains
\bea
&&\d_2{\cal{V}}_P^{(a)}+\d_1{\cal{V}}_P^{(b)}\rightarrow
-\frac{i}{4}\int d^2\theta~d^2\bar{\theta}~
\frac{d^4p_0~d^4p_1~d^4p_2~d^4p_3~d^4p_4}{(2\p)^{20}}~
\d(\sum_{i=1}^4 p_i)~\nonumber\\
&&~~~~~~\nonumber\\
&&~~~~~~~~
\int \frac{d^4k}{(k+p_1)^2 k^2 (k-p_4)^2
(k+p_1+p_2)^2}\left[-(k-p_0)^2~\frac{e^{-\frac{i}{2} k\times p_0}~
e^{\frac{i}{2}(k-p_4)\times (p_4-p_0)}}{(k-p_0)^2}\right.\nonumber\\
&&~~~~~\nonumber\\
&&~~~~~~~~~~~~~~~~~~~~~~~~~~~~~~~~~~~~~~~~~~~~\left.+
(k-p_4+p_0)^2~ \frac{e^{-\frac{i}{2}(k-p_4)\times
p_0}~e^{-\frac{i}{2}(p_4-p_0)\times k}}{(k-p_4+p_0)^2 }\right]
\nonumber\\
&&~~~~~~\nonumber\\
&&~~~~~~~~~~~~~~~~~~~~~~~~
{\rm Tr}[K(p_0){\bf W}^\a(p_1){\bf W}_\a(p_2)]~
{\rm Tr}[\bar{{\bf W}}^\ad(p_3)\bar{{\bf W}}_\ad(p_4-p_0)]\nonumber\\
&&~~~~~~\nonumber\\
&&~~~~~~~~~~~~~~~~~=0
\label{cancel}
\eea
Continuing the calculation, one finds
\bea
&&\d_2{\cal{V}}_P^{(b)}+\d_1{\cal{V}}_P^{(c)}\rightarrow 0\nonumber\\
&&~~~~~\nonumber\\
&&\d_2{\cal{V}}_P^{(c)}+\d_1{\cal{V}}_P^{(d)}\rightarrow \d_1
\G_{PT}^{(2)}
+\d_2 \G_{PT}^{(2)}\nonumber\\
&&~~~~\nonumber\\
&&\d_2{\cal{V}}_P^{(d)}\rightarrow \d_2 \G_{PT}^{(1)}
\label{dP}
\eea
In this way we have shown that the terms produced from the variation
of ${\bf W}^\a(p_1)$ and ${\bf W}_\a(p_2)$ in (\ref{varexample}) are
completely cancelled by the variation of the untwisted terms
from the ${\cal V}$ vertex in the five-point function. Following
exactly the same pattern the twisted ${\cal V}$ vertex produces a
variation that compensates the variation from $\bar{{\bf W}}^\ad(p_3)$
and $\bar{{\bf W}}_\ad(p_4)$ in the four-point contribution in
(\ref{varexample}).

\vskip 34pt
\noindent
\begin{minipage}{\textwidth}
\begin{center}
\includegraphics[width=0.70\textwidth]{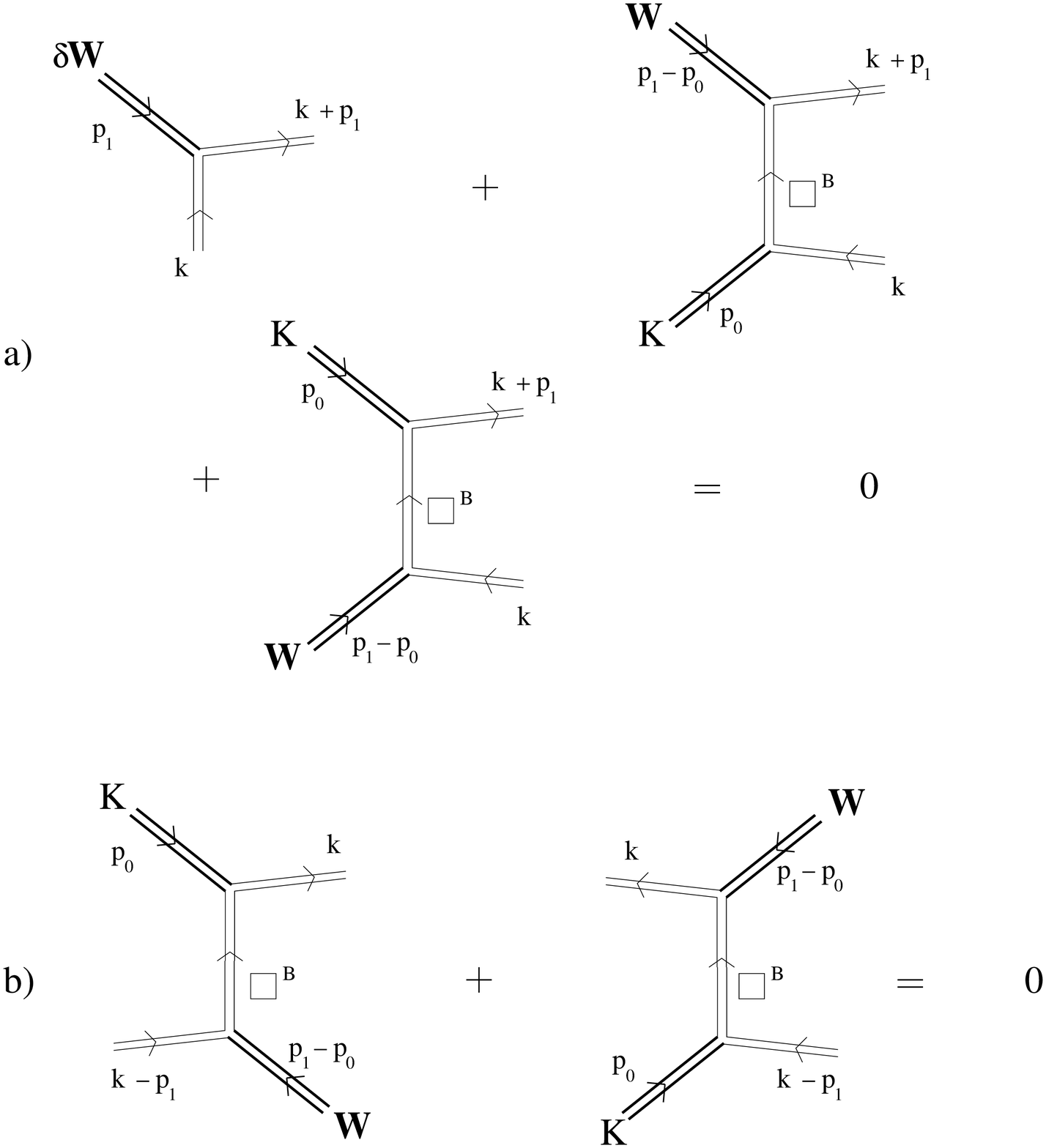}
\end{center}
\begin{center}
{\small{Figure 4:
general cancellation mechanism}}
\end{center}
\end{minipage}

\vskip 34pt
We can implement the cancellation graphically as in
Fig. $4$, which shows the interplay between twisted and untwisted vertices.
In fact the previous
argument can be generalized and
the gauge invariance of the complete one-loop
effective action can be obtained. The reasoning is the following:
one rewrites the quadratic quantum Lagrangian in (\ref{oneloopaction})
in a manifestly background gauge invariant way
\beq
 -\frac{1}{2 g^2} {\rm Tr} \left( V * \Box^B * V -
i V * \{ {\bf W}^{\alpha},\nabla_\alpha V\}_* -
i V * \{ {\bf {\bar{W}}}^\ad,\bar{\nabla}_\ad V\}_*\right)
\label{actinv}
\end{equation}
with the definition $ \Box^{B}\equiv \frac 12
\nabla_a^B \nabla^{Ba}$. As explained in \cite{GZ} one can perform
one-loop calculations using background covariant $D$-algebra and
background covariant propagators. In particular,
making a background gauge transformation on the background fields
only, the first term of the action in (\ref{actinv})
gives the variation
in (\ref{5var}) with $\Box$ replaced by $\Box^B$, as shown in Fig. $4b$.
The terms containing ${\bf W}$ and ${\bf {\bar{W}}}$ are
represented graphically in Fig. $4a$, with covariant propagators
 $1/\Box^B$ . In this way, using the previous argument
one establishes that the one-loop effective action is indeed
background gauge invariant.

\vspace{0.8cm}

We conclude with the following remarks. We have checked that the
noncommutative ${\cal N}=4$ supersymmetric Yang-Mills theory can be
treated perturbatively in a consistent and technically viable
framework. Thus one can continue in the program and start computing
correlation functions in perturbation theory. In \cite{GHI,IKK}
gauge invariant operators have been constructed in the noncommutative
theory through the introduction of appropriate Wilson lines
\beq
{\cal W}(x,C)=P_*\exp\left(i\int_0^1
d\s\frac{d\zeta^\m}{d\s}A_\m(x+\zeta(\s))\right)
\label{wilson}
\eeq
where $P_*$ denotes generalized path ordering in terms of the $*$-product and
$C$ is a straight line. The calculations that have been presented so
far are in a component approach. We can extend them to
supersymmetric theories in a full superspace formulation.
Indeed the Wilson line can be represented in superfield language as
\beq
{\cal W}(x,C)=P_*\exp\left(i\int_0^1
d\s\frac{d\zeta^{\a\ad}}{d\s}\G_{\a\ad}(x+\zeta(\s))\right)
\label{superwilson}
\eeq
It can be used efficiently in perturbative calculations of
correlation functions of gauge invariant operators.

The Wilson lines introduced in \cite{GHI,IKK} have been used to obtain a
gauge-completion of the one-loop four-point function \cite{HL}.
In the background field method approach that we have applied to
derive the one-loop effective action, Wilson lines naturally
appear as follows:
\beq
\nabla^B_a= {\cal W}*\pa_a~{\cal W}^{-1}\qquad\qquad\quad
\frac{1}{\Box_B}={\cal W}*\frac{1}{\Box}*{\cal W}^{-1}
\eeq
where ${\cal W}$ is defined in (\ref{superwilson}).
In this way contributions to the one-loop effective action
can be rewritten as
\bea
&&{\bf W^\a}*\frac{1}{\Box_B}*\nabla^\b_B \bar{\nabla}^\bd_B \bar{{\bf W}}^\ad
*\frac{1}{\Box_B}\dots
\rightarrow {\bf W^\a}*{\cal W}*\frac{1}{\Box}*{\cal W}^{-1}*
\nabla^\b_B \bar{\nabla}^\bd_B \bar{{\bf W}}^\ad *{\cal W}
*\frac{1}{\Box}\dots\nonumber\\
&&~~~~~
\eea

\vspace{1.5cm}

\noindent
{\bf Acknowledgements}

\noindent
We acknowledge stimulating discussions with Ruggero Ferrari and
Soo-Jong Rey.
This work has been partially supported by INFN, MURST and the
European Commission RTN program HPRN-CT-2000-00113 in which M.P., A.S. and D.Z.
are associated to the University of Torino.

\end{document}